\documentclass[aps,pre,preprint,groupedaddress,showpacs]{revtex4}
\usepackage{bm}
\usepackage{times}
\usepackage{amsmath}
\usepackage{amsfonts,amssymb}

\usepackage{verbatim}
\begin{document}

\title{Kinetic description of avalanching systems}

\author{M. Gedalin$^1$, M. Balikhin$^2$, D. Coca$^2$, G. Consolini$^3$, and R. A. Treumann$^4$}
\affiliation{$^1$Ben-Gurion University, Beer-Sheva, Israel\\
$^2$ACSE, University of Sheffield, Sheffield, UK\\
$^3$Ist. Fisica  Spazio Interplanetario, Istituto Nazionale di Astrofisica, Rome, Italy\\
$^4$Max-Planck-Institute for Extraterrestrial Physics, Garching,
Germany}

\begin{abstract}
Avalanching systems  are treated analytically using the
renormalization group (in the self-organized-criticality regime)
or mean-field approximation, respectively. The latter describes
the state in terms of  the mean number of active and passive
sites, without addressing the inhomogeneity in their distribution.
This paper goes one step further by proposing a kinetic
description of avalanching systems making use of the distribution
function for clusters of active sites. We illustrate application of the kinetic formalism to a model  proposed for the description of the avalanching processes in the reconnecting current sheet of the Earth magnetosphere.  
\end{abstract}
 \maketitle

\section{Introduction}
Many natural systems, which work in an open configuration, respond to external disturbances showing scale-invariant discrete events \citep{Set2001}. One common feature of these systems is the development of a local threshold instability in an avalanching manner. In the late 80s the concept of self-organized criticality (SOC) was proposed by Bak et al. \citep{Bak87PRL} for the dynamical-statistical behavior of such systems. 
 SOC has been applied
to a variety of systems (see \citet{Jensen1998} and references
therein for a list of some such systems). Although SOC can exist,
strictly speaking, only in the limit of infinitely slow external
input where complete separation of time scales is achieved \citep{HK92} , it has also been applied to presumably
avalanching systems with strong driving. A good example of such
systems is space plasma and, in particular, the plasma in Earth's
magnetotail under magnetic substorm conditions \citep{Chang99}.
Since SOC is questionable for such strongly driven systems we, in
what follows, address to them as to avalanching systems, bearing
in mind the avalanche-like propagation of local instabilities. Up
to date the most often used tool for studies of such systems is
numerical modeling. The usual analytical approaches proposed so far are the renormalization group methods (see,e.g., Refs.~\citep{Jensen1998,renorm} and references therein), and the mean-field description (see,e.g., Refs.~\citep{Jensen1998,mean} and references therein). The renormalization group methods assume scaling from the very beginning and are applied only in the close vicinity of the stationary (critical) point, that is, in the self-organized criticality regime.    The mean-field approach is based on the analysis of the mean
number of active, passive, and critical sites. It is not
restricted to the criticality range only including it as the limit
of zero number of active sites. Mean-field approximations predict self-organized criticality in the limit of zero average number of active sites and, strictly speaking, are applicable only for system dimension exceeding some critical number, often well above the dimension of real physical systems \citep{bs}. Mean-field obtained exponents are often consistent with those found experimentally and numerically for lower dimensions too but no quantitative explanation is given. On the other hand, deviations from these exponents for real systems are quite usual.  The mean-field approach does not take into
account the tendency of the active sites to organize in clusters.
Indeed, if avalanches of various durations and sizes are present,
the distribution of active sites at any moment should be very
inhomogeneous. In the present paper we propose a novel approach to
the analytical description of avalanching systems which is based
on the kinetic equation for the distribution function for active
site clusters. We demonstrate the power of the kinetic formalism applying to the model which was recently proposed as a model of avalanching reconnection in the current sheet of the Earth magnetosphere \citep{Gedalin2005}.

\section{Kinetic equations for clusters}
The mean-field approach has the obvious drawback of ignoring that
active sites have the tendency to appear in clusters. These
clusters are, in fact, the instantaneous snapshot of the
developing avalanches, so that the size of each cluster is time
dependent, $w=w(t)$. However, when considering many coexisting
clusters, we may describe their behavior with the help of the
distribution function $f(w,t)=dN/dw$, where now the cluster size
$w$ and time $t$ are independent variables. The evolution of the
single cluster size will be translated into the evolution of the
distribution function. The total number of active sites is given
by the integral
\begin{equation}
N_a=\int_0^\infty wf(w)dw
\end{equation}
We have to introduce also the number of passive sites $N(0)$
(similar to what is done in case of a Bose-gas, where the number
of particles in the lowest state is macroscopically large). Then
$N_a+N(0)=\text{const}$.

Let $P_+(w_1,w_2)$ be the probability of the cluster growth (per
unit time), and $P_-(w_1,w_2)$ be the probability of shrinking.
Then
\begin{equation}\label{eq:kinetic1}
\begin{split}
\frac{\partial f(w)}{\partial t}&= \int_0^w P_+(w,w') f(w')dw' +
\int_w^\infty P_-(w,w')f(w')dw'\\
&-\int_w^\infty P_+(w',w)f(w)dw' - \int_0^wP_-(w',w)f(w)dw'+
\gamma(w)N(0)-\frac{f(w)}{\tau(w)}
\end{split}
\end{equation}
The  term  $\gamma(w)N(0)$ in \eqref{eq:kinetic1} describes the
birth of active states due to external driving, while the last
term takes into account the finite life-time of clusters, i.e. the
transition to the passive state (Bose-Einstein condensation). If
the driving is sufficiently strong and avalanche merging is not negligible, the
kinetic equation \eqref{eq:kinetic1} should be completed with the
with the time-dependent ``nonlinear" merging terms
\begin{equation}\label{eq:merging}
\begin{split}
\left(\frac{\partial f(w)}{\partial t}\right)_{m}&=\int P_1(w,w_1,w_2)
f(w_1)f(w_2) \delta(w-w_1-w_2)dw_1 dw_2 \\
& - \int P_1(w_1,w,w_2) f(w)f(w_2)\delta(w_1-w-w_2)dw_1dw_2.
\end{split}
\end{equation}
Merging becomes progressively more important when the average fractional density of active sites increases. When this density  is not too large (it does not have to be small though, in contrast with the SOC regime), merging will be still relatively weak and can be further studied perturbatively. Strong merging corresponds to the very strong driving, so that the system behavior is, at least partially, forced externally. In the present paper we assume that  driving is moderate (not weak and not exceptionally strong) so that 
merging can be ignored at this stage, deferring treatment of very strongly driven
systems to elsewhere. In our case one can expect that there is a wide range (inertial interval) of cluster sizes in which the distribution shape is  independent of the external  driving and is determined by internal dynamics and/or space dimension. 

In general, the distribution function $f(w)$ would depend on the growth and shrinking
probabilities. We shall consider here the class of systems where growth
and shrinking occur only at the boundaries of clusters. It should be noted that the dynamics inside clusters may induce transitions between active and passive sites, producing, e.g. "punctuated" clusters for the  classical sandpile model \citep{Bak87PRL}, where an active site becomes passive at the next step. We shall measure the size of such cluster including the passive (receiving) sites as well, so that the internal dynamics does not affect the cluster size. Situation may be more complicated when clusters are developed fractals, with  tunnels
appearing and crossing the cluster \citep{Set2001}. Such systems would probably require special treatment. We restrict ourselves here with the clusters which grow of shrink at their boundaries. Space and laboratory plasma systems \cite{plasma} seem to belong to this class. 

In this
case the probabilities are nonzero only for $|w'-w|=\Delta\ll w$,
so that \eqref{eq:kinetic1} can be written
\begin{equation}\label{eq:kinetic1a}
\begin{split}
\frac{\partial f}{\partial t}& = - \tilde{P}_-(w)\sigma(w) f(w) -
\tilde{P}_+(w)\sigma(w) f(w)\\
& + \tilde{P}_-(w+\Delta)\sigma(w+\Delta)f(w+\Delta)
+ \tilde{P}_+(w-\Delta)\sigma(w-\Delta)f(w-\Delta),
\end{split}
\end{equation}
where $\sigma(w)$ is the density of states. This approximation is not valid for small $w$, where the cluster kinetics should be strongly affected directly by driving. We seek for an approximate description of the cluster kinetics in the range where it is determined but the internal features of the system rather than by external influence. It is obvious, that if a large size strong driving is applied the reaction of the system would be a forced reaction and not  self-organized in any way.  

The approximation may be not accurate for largest clusters either, since possible fractality \citep{Set2001} of clusters may result in the breakdown of independence of probabilities at neighboring active boundary sites. Indeed, all numerical simulations \citep{Jensen1998} show distortions for very small and very large $w$. Thus, the physical sense of our approximation is that we are working in the \textit{inertial interval} far from both limits. According to existing analyses, such interval exists almost always.

For one-dimensional
clusters $\sigma(w)=1$ or $\sigma(w)=2$ (the latter holds for
growth in both directions). This allows immediate $n$-dimensional generalization. 
Let $w$ be a linear measure of a  cluster (effective \textit{radius}), and let $D$  be the cluster volume. The density of states $\sigma$ is then the cluster surface area. In general, 
$D\propto w^\mu$, $\sigma\propto w^\nu$, $n\geq \mu >\nu \geq n-1$, where $\mu$ and $\nu$ are fractal dimensions of the cluster volume and boundary, respectively. 
 Taylor expanding
\eqref{eq:kinetic1a} we arrive at the following differential
equation
\begin{equation}
\frac{\partial f}{\partial t}=\frac{\partial }{\partial w} (\alpha \sigma f) +
\frac{\partial^2 }{\partial w^2} (\beta \sigma f),
\label{eq:kinetic1b}
\end{equation}
where $\alpha=\Delta(\tilde{P}_- - \tilde{P}_+)$ and
$\beta=(\Delta^2/2) (\tilde{P}_- +\tilde{P}_+)$. The stationary
solution
\begin{equation}
f\propto (1/\beta\sigma) \exp\left[-\int (\alpha/\beta) dw\right],
\label{eq:kineticstat}
\end{equation}
exists only if $\alpha>0$. In general, $\alpha$ and $\beta$ can depend on $w$. Both describe the local growth (shrinking)  per site at the cluster surface. Their dependence on the cluster size would mean essentially that the growth and shrinking probabilities as well as variation of the affected neighbor zone at some site depend on what happens at other sites. While,  in principle, this cannot be excluded (waves could transfer information across the cluster or long range forces are involved \citep{Set2001}), many  avalanche systems seem to be governed by local dynamics, so that it is natural to consider (at least at this stage) the case of   
 probabilities independent of $w$ (see, however, comment in section~\ref{sec:burning}). 
One finds
\begin{equation}
f\propto \sigma^{-1} \exp (-w/w_c)=w^{-\nu}\exp(-w/w_c), \label{eq:wc}
\end{equation}
with $w_c=\text{const}$. The obtained  $f=dN/dw$ describes the distribution of linear sizes (effective radii). For the distribution of the cluster volumes one has
\begin{equation}
\frac{dN}{dD}=\frac{dw}{dD}\cdot\frac{dN}{dw}\propto D^{(1-\mu-\nu)/\mu} \exp\left(-
AD^\frac{1}{\mu}\right). \label{eq:dndd}
\end{equation}
In the  mean-field limit $n\gg 1$ \citep{mean} one has $(dN/dD)\propto D^{-2}$. 

The derived expressions assume isotropy. If the system is anisotropic and/or a preferential shape of cluster exists, e.g. clusters are elongated \citep{Set2001}, the above treatment may have to be modified by considering vector $\bm{w}$ describing linear sizes along principal axes.  These modifications are of technical character and do not change substantially the basic equations and conclusions. Yet, they require a more lengthy analysis and cannot be presented in a letter. We will provide this analysis elsewhere. 
 
\section{Burning model}\label{sec:burning}
The above theory can be illustrated on the simple "burning" model \citep{Gedalin2005}
described below. In this model each site is characterized by its
temperature, $T(x)$. The external driving is random heating of the
sites. The amount $q$ of heat per unit time is going to a site
with probability $p$, so that the average heat transfer from
outside (in driving) is $pq$.  The temperature of a passive site
(the one which is not burning) changes according to
\begin{equation}
\frac{dT}{dt}=qp(1-\eta(t)), \label{eq:t}
\end{equation}
where $\eta(t)$ is a random number, $|\eta|\leq 1$, so that
$\langle \eta(t)\eta(t')\rangle=\delta(t-t')$. Once $T>T_c$, where
$T_c$ is some critical temperature, the site becomes active. An
active site burns and produces heat at the rate $J=\mu T$,
$\mu<1$. During the burning stage the temperature decreases
(unless driving is strong enough to force permanent burning). When
the temperature drops below some value $T_l$ such that
$T<T_l<T_c$, the burning ceases and the site becomes passive
again. Part of the heat release is lost (radiated away), while the
other part, $a J$, is transferred (isotropically) to the closest
neighbors. Summarizing the above, the heat release can be written
approximately as
\begin{equation}
J=\mu T\theta(T_c-T)\theta(-dT/dt)\theta(T-T_l) + \mu T\theta(T-T_c)
\label{eq:j}
\end{equation}
where $\theta(x)$ is the step function. The term $\theta(-dT/dt)$
is an approximate manifestation of the history dependent
(hysteresis) burning for $T_c>T>T_l$ (burning now if it was burning
at the previous moment/step and not burning otherwise). This
expression is not quite correct for the temperature of a site does
not have to change monotonically when an avalanche develops. We
leave the more detailed discussion of this for another paper
especially devoted to this model. For the purposes of the present
discussion such details are irrelevant, and we consider
\eqref{eq:j} as a sufficiently precise description of the burning
process. If an active site would be left alone, its temperature
would decrease as $T=T(0)\exp(-\mu t)$. Here the quantity
$\tau\approx (1/\mu) \ln(T_c/T_l)$ has the meaning of the life
time of an active site if it were not affected neither by other
sites nor external driving. Let $\Delta_t$ be the time step and
$\Delta_l$ the site size.  The amount of heat a site $x$ is
receives is given by
 \begin{equation}
\frac{dT(x)}{dt}=qp[1-\eta(t)] + a [J(x+\Delta_l)+J(x-\Delta_l)]
-J(x), \label{eq:dt1}
\end{equation}
which we write in the following form
\begin{equation}
\frac{dT(x)}{dt}=qp[1-\eta(t)] + (2a -1)J +
\frac{a\Delta_l^2}{2} \frac{\partial^2 J}{\partial x^2}.
\label{eq:dt2}
\end{equation}
Integrating \eqref{eq:dt1} over a cluster of the size $w$, one gets
\begin{equation}
\frac{d}{dt} \int Tdx= qpw + (2a-1)\int Jdx - J_{b},
\label{eq:int}
\end{equation}
where we averaged over time the random fluctuations of the input
$\eta$. The last term is the heat flux at the boundaries.

The probability of growth should be proportional to the heat flux
from the  active site at the cluster boundary to the neighboring
passive sites. This probability should depend on the temperature
of the passive sites. In the stationary regime the time-average
growth probability would be determined by the average temperature
$T_p$ of passive sites. Thus, growth is essentially independent of
the cluster size. Respectively, the shrinking probability depends
on the state of the boundary site and is not particularly
sensitive to the cluster size either. In this case the parameters
$\alpha$ and $\beta$ are constant, and one expects that the
cluster distribution is an exponential, $f\propto \exp(-w/w_0)$.
However, if the heat transfer in the active area is suppressed
(active sites do not easily accept heat from active neighbors)
spreading from the central regions with the constant speed up to
the cluster boundaries, one estimates that $P_+\propto 1/w$, while
$P_-\approx \text{const}$. In this case
$\alpha=\Delta(a_1-a_2/w)$, and $\beta=(a_1+a_2/w)\Delta^2/2$, and
$
f\propto (w+w_0)^\lambda \exp(-w/w_c),\label{eq:c}
$
where $w_0$, $w_c$ and $\lambda$ are constants. In the range
$w_0\ll w\ll w_c$ (if such this range does exists at all) a
power-law distribution should be observed.
In the opposite case, when the heat is transferred immediately
from the inside to the cluster boundaries, $P_+\propto w$ and
$P_-\approx \text{const}$, no stationary state can exist, since
$\alpha<0$ for sufficiently large $w$. Such systems are unstable
and are disrupted into avalanches which will cover the entire
system.

\section{Conclusions}

We proposed a kinetic approach to the description of avalanching
systems, defining a distribution function $f(w,t)$ for the
clusters of active sites. In this way we derived a kinetic
equation for the temporal evolution of $f(w,t)$ and analyzed its
steady state limit in the inertial range, sufficiently far from the smallest scales where driving explicitly shows up, and sufficiently far from the larges scales where fractality and merging become progressively more important. The stationary distribution function $f(w)$
depends, in general, on the probability of the micro-processes
resulting in cluster growth and shrinking, that is, the processes
governing the evolution of avalanches. In the case of locally
induced growth at the boundaries the shape of the distribution is
determined by the dimension of the system (or fractal dimensions of  clusters if they are not compact). There is no
sensitivity to the input details. The obtained universal shape of the distributions is not limited to the weak driving regime or to the system dimension above some critical value, and can be used for direct and easy comparison with experiments and numerical modelling.  The total average driving should
affect the state of the system, as we have shown in a particular
model. The estimates given in the present model represent just the
first step toward a more elaborated kinetic model of the dynamics
of avalanches. 
We remark that our analytical predictions have been checked by 1D and 2D burning model simulations to be reported elsewhere.

\begin{acknowledgments}
This work has been performed within the framework of the
``Observable features of avalanching systems" ISSI Science Team.
\end{acknowledgments}

\end{document}